\documentclass[aps,prl,preprint,superscriptaddress]{revtex4}
\usepackage{graphicx}

\newcommand{\tq}{TTF-TCNQ }
\newcommand{\q}{TCNQ }
\newcommand{\f}{TTF }
\begin{document}

\title{Scanning Tunneling Microscopy in \tq :direct proof of phase and amplitude
modulated charge density waves } 

\author{Z. Z. Wang}\author{J. C. Girard} 
\affiliation{Laboratoire de Photonique et de Nanostructures, CNRS,
route de Nozay,
91460, Marcoussis, France}
\author{C. Pasquier}\author{D. J\'erome}
\affiliation{Laboratoire de Physique des Solides (CNRS), Universit\'e
Paris-Sud, 91405, Orsay, France}
\author{K. Bechgaard}
\affiliation{Polymer Science Department, Research Center Ris\oe, 
DK 4000, Roskilde,
Denmark} 

\date{\today}

\begin{abstract} Charge density waves (CDW) have been studied at the surface of a
cleaved  \tq single crystal  using a low temperature scanning tunneling microscope (STM)
under ultra high vacuum (UHV) 
conditions, between 300K and 33K with molecular resolution. All
 CDW phase transitions of
\tq have been identified. The measurement of the modulation wave vector 
along the
\textbf{a} direction provides the first evidence for the existence of domains
comprising single plane wave modulated structures in the temperature regime where the
transverse wave vector of the CDW is temperature dependent,
as hinted by the theory more
than 20 years ago.
\end{abstract}

\maketitle

%\section{Introduction}
The discovery of the  organic molecular crystal
tetrathiafulvalene-tetracyanoquinodimethane 
(\tq), comprising weakly coupled one dimensional (1D) molecular stacks,
created a tremendous turmoil in 1973
\cite{Heeger,Cowan}. This was the first molecular crystal to show a conductivity
approaching that of conventional metals at room temperature and exhibiting a metal-like
behaviour on cooling. 
A
partial charge is transfered from  \f to \q stacks and the charge density $\rho$
potentially available for transport is determined by the value of $k_F$ at which
the bonding
\q band crosses the antibonding \f band, leading to $2k_{F}$=$\rho\pi/b$ in the 1D
band picture, where \textit{b} is the unit cell length .
Between 54K and 38K, CDW's successively develop in the \q and \f
stacks. These transitions have been ascribed to the instability of a one
dimensional electronic gas due to the Peierls mechanism, \textit{see reference}
\cite{Schulz} for a review.

 When the CDW's are
active on both kinds of stacks frustration arises and the 2D ordered superlattice can be
described by plane waves with the wave vectors 
\textbf{$ q_{+}$} = [+$q_{a}(T)$, $2k_F$] or \textbf{$q_{-}$}
=[$-q_{a}(T)$, $2k_F$]. Both lead to  configurations which are energetically
equivalent. The wave vector  $ q_{+}$ gives rise to  a charge
modulation such as
$\rho (r)$ = $\rho_{+}$cos\textbf{$(q_{+}r +\theta_{+})$} which is  a CDW
of fixed amplitude and a phase varying like $q_{a}a$ along the $a$ direction,
 and
similarly for the $q_{-}$ wave vector. Consequently, the diffraction pattern of
the CDW state should display  an equal number of domains characterized by the
vectors
\textbf{$q_{+}$} and
\textbf{$q_{-}$} . 

There also exists another possibility, namely: the
superposition of the two  plane waves 
\textbf{$q_{+}$} and \textbf{$q_{-}$} which leads to a CDW with constant
phase but a modulated amplitude  along the $a$ direction \cite{Bjelis,Abrahams},
double-$q$ configuration. The only solution which can take advantage of the
commensurability energy related to the transverse commensurate periodicity through the
fourth order Umklapp term in a Landau-Ginzburg expansion  is the double-$q$ configuration
\cite{Abrahams}. This means that both wave vectors are simultaneously activated
below 38 K with four satellite spots at \textbf{$\pm q_{+}$} and
\textbf{$\pm q_{-}$} in the reciprocal space around a main Bragg spot. On the other
hand, it has been pointed out that the phase modulated solution should be the most
stable one in the incommensurate transverse wave vector temperature regime and also the
only one to provide a smooth onset at 49K
\cite{Abrahams}. The presence of a microscopic coexistence of vectors
\textbf{$q_{+}$} and \textbf{$q_{-}$} below 38K has been shown by  X-ray
diffuse scattering \cite{Kagoshima} and a structural determination \cite{Bouveret}.
However, in spite of the data of an early STM study of \tq \cite{Nishiguchi} showing a phase
modulated 2D structure at 42 K  there exists  no direct  evidence  of a transition from a phase
modulated regime between  49 and 38 K 
where the temperature dependence makes the amplitude of
$q_a$  sliding to an amplitude modulated situation below 38 K. Diffraction experiments performed
on a bulk sample have failed to provide a clue since they 
cannot tell the difference between an amplitude modulated configuration and one
in which the phase is modulated  with an equal number of domains with
\textbf{$q_{+}$} and
\textbf{$q_{-}$}. Therefore, only those specific techniques like STM probing the
sample locally are likely to provide an answer to this problem.

The present STM investigation of a \tq single crystal has been performed in
a broad temperature range (33-300 K). The primary goal was to achieve the best
possible experimental conditions in order to provide local information regarding
the development of 3D ordered CDW's below 54K. This work brings the first direct
experimental proof for the existence of phase modulated and amplitude modulated
CDW's between 49-38 K and below 38 K respectively and also supports the model
proposed by theoreticians more than 20 years ago
\cite{Bjelis,Bak}.

%{Experimental Conditions}

The experiment was carried out in an UHV-LT-STM system with separate UHV chambers
for STM measurements, sample and tip preparation. The base pressure in each chamber
is in the range of $10^{-11}$ mbar.  A commercially available LT-STM head is used in
this study and the entire scanning unit (including tip, sample, piezo tube, piezo
motor and damping system) is inserted in a thermostat with four  gold-plated cold
shields (Omicron LT-STM). The sample temperature is controlled by  a Lake Shore DRC
91C controller. Typical temperature fluctuations are less than 20 mK in 200 seconds
with an  average temperature drift below 50 mK per hour. Mechanically sharpened
Pt/Ir tips were used. The durability of the tips has been demonstrated
by their ability to get molecular resolution of TTF-TCNQ for hours. The quality of
the  tips is  checked by their ability to obtain atomic resolution on a gold surface
before and after imaging of \tq.
We image the sample using a constant current mode. The maximum data rate is 100 KHz
and the typical time needed to record one image is 200 seconds. 

Crystals of TTF-TCNQ with nice looking natural faces and typical 
dimensions of 3 $\times$ 0.5 $\times$ 0.05 mm$^3$ are 
selected for this experiment. A clean (001) surface is obtained by 
cleaving the single crystal with a razor blade in air just before insertion. Direct
exposure to air is restricted to less than 2 minutes. In order to avoid micro-cracks in
TTF-TCNQ while cooling or warming, the temperature variation rate is kept at 1 K per
minute.

%\section{Results}
%%%%figure1%%%%%

\begin{figure}
\includegraphics[width=5.5cm]{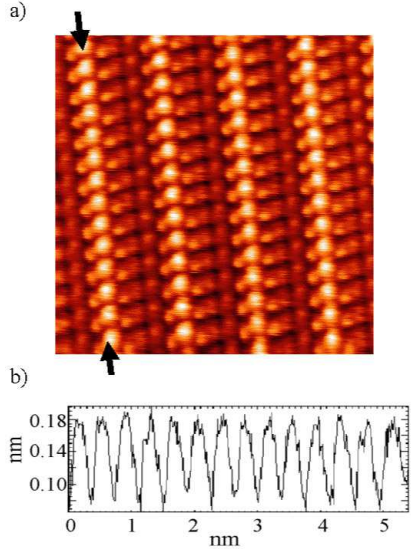}
\caption{a) STM image of the \textbf{a-b} plane of TTF-TCNQ taken at 63K.The
image area is 5.3 nm $\times$ 5.3 nm. b) shows the profile along a TCNQ
stack indicated by black arrows.}
\label{fig.1}
\end{figure}

 Figure 1a displays a typical
image of the \textbf{a-b} plane (area 5.3 $\times$ 5.3 nm$^2$) 
obtained in a constant current mode
(I=1 nA, V=50 mV) at 63 K where a 1D structure of parallel chains is clearly
observed with one set of chains containing a triplet of balls and the other
 a doublet.
 According to the calibration of the piezo at low temperature, the distance between
similar chains is 1.22 nm and 0.38 nm between units along the chain direction,
see fig.1b. Both distances compare very well with the \textbf{a} and
\textbf{b} lattice constants, \textit{b} = 0.3819 nm and \textit{a} = 1.229 nm
\cite{Kistenmacher}.
We can ascribe the triplet 
feature in fig.1a to the \q in agreement with the early work of Sleator and
Tycko \cite{Sleator}.  The \f molecule appears usually as a single ball feature in STM
imaging although reports of doublet structures have also been made in the
literature \cite{Kato}. An extensive interpretation of the \tq image in the
absence of CDW will be given in a forthcoming paper\cite{Ordejon}and the present work 
is restricted to the physics provided by STM images of the \q molecules only.
 No bias voltage dependence  (polarity) of the
image was observed during our measurements in agreeement with the expected
conducting nature of the surface
\cite{Sacks}. 
In the whole temperature domain where the sample is metallic
\textit{i.e.} above 54K, images like  fig.1a were observed and  we could not detect
any modulation on the STM image besides that provided by the uniform \tq lattice.
 Therefore, the periodic modulation along the \q stacks reported in
ref \cite{Nishiguchi} at 61K could be related to static CDW's stabilized by defects
or steps on the surface as noticed by the authors.  

Below 54K a two dimensional superstructure
restricted to the
\q chains with a period of 2a $\times$ 3.3b appears in the image (see fig.2a).

%%%%%%figure2%%%%%%

\begin{figure}
\includegraphics[width=8.8cm]{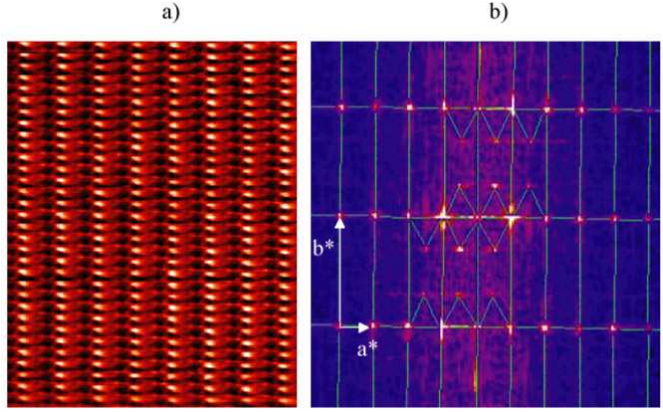}
\caption{a) STM image of the \textbf{a-b} plane of TTF-TCNQ taken at 49.2K. The image
area is 8.7 nm x 11.9 nm, b) Fourier transformed pattern showing the
2a$\times$3.39b CDW ordering.}
\label{fig.2}
\end{figure}

The modulation wavevector (shown in fig.2b by Fourier transforming the image)
does not vary
down to 49K. On further cooling, the transverse modulation vector becomes
incommensurate (IC) and a temperature dependence  $q_{a}(T)$  is observed
without noticeable change along \textbf{b}, figs.3a,b.
The Fourier
transformed image shows that the modulation can be described by a single wave vector
\textbf{$q_+$} or \textbf{$q_-$} in the temperature domain 49-38 K. 
However, a  transverse commensurability ($\times 4$) 
arises abruptly at 38 K. The
ordering of the charge density modulations both along \textbf{a} and \textbf{b} directions at
36.5 K is  presented in figs.3c,d. 
Below 38K (low temperature commensurate
phase) a double-\textbf{$q$} CDW modulation \textbf{$q_+$} and
\textbf{$q_-$} is identified.

%%%%figure3%%%%

\begin{figure*}
\includegraphics[width=12.5cm]{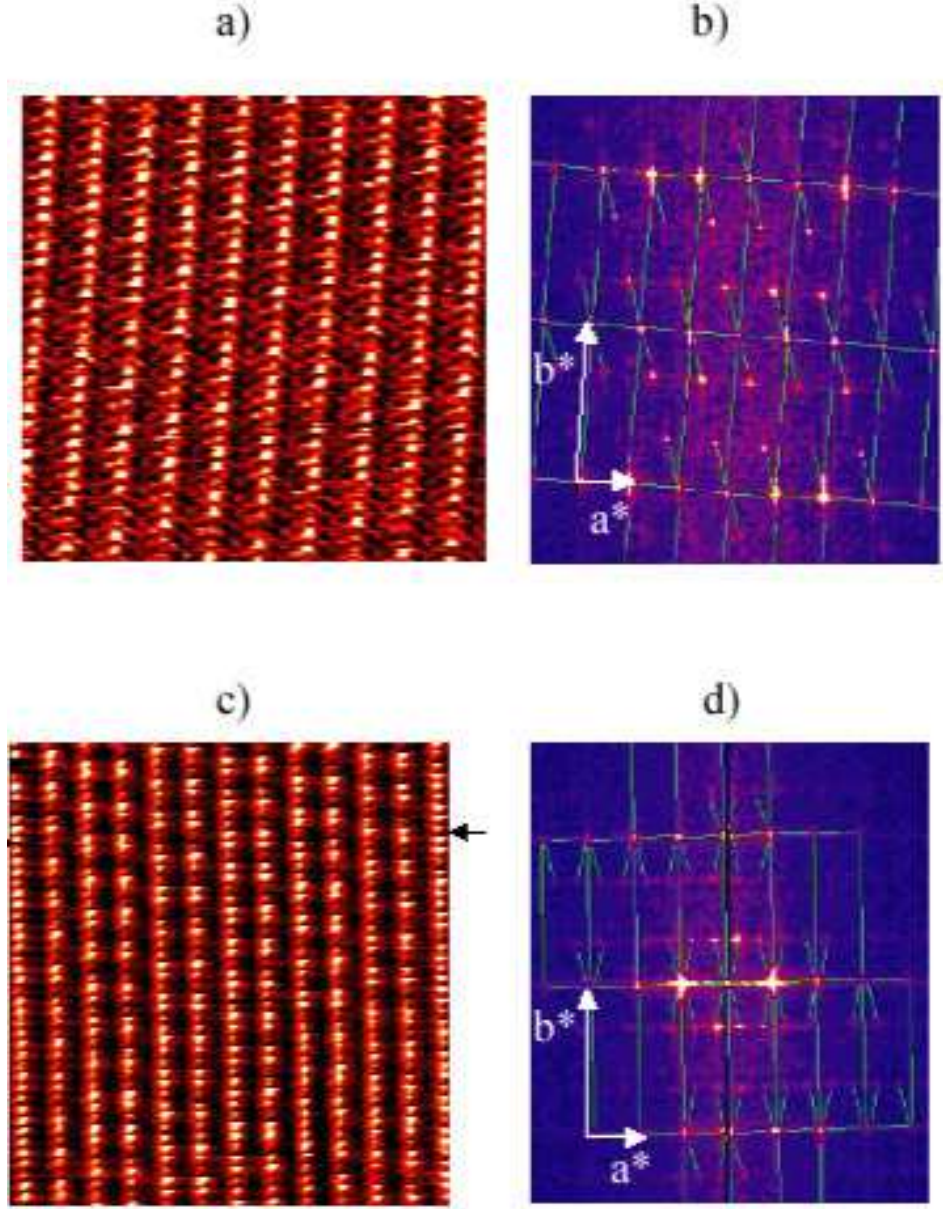}
\caption{a) STM image of the \textbf{a-b} plane of TTF-TCNQ taken at
39K, the image area is 9.3 nm x 6.9 nm, b) Fourier transformed pattern
showing the single-q CDW in the sliding temperature domain, c) STM image
of the
\textbf{a-b} plane of TTF-TCNQ taken at 36.5K, the image area is
14.8$\times$14.7 nm, d) Fourier transformed pattern showing the
double-q (4a$\times$3.3b) CDW in the commensurate phase.}
\label{fig.3}
\end{figure*}

The images presented above are the first to report a study of  the 2D superlattice
structure of
\tq in real space below the Peierls transition down to the temperature of 33 K. The value and
temperature dependence of the modulation wave vector  are in very good agreement with the
detailed X-ray
\cite{Pouget,Khanna,Kagoshima} and neutron scattering \cite{Ellenson,Ellenson77}
reports (see fig.4a).

We can provide a real space signature of the intermediate temperature regime
in which the transverse period is evolving with temperature (the sliding regime)
before a lock-in takes place at 38 K. Although the signal coming from the CDW
modulation is always dominant in all our scans (with a corrugation of 0.21 nm at
36.5 K along the
\q stacks) it does not  overcome the corrugation coming from the underlying
\q lattice, namely 0.12 nm. Thanks to the coexistence between CDW and original
lattices on the images, molecular resolution can be obtained in the CDW condensed
state at low temperature.

 This is
\textit{at variance} with layered compounds such as 1T-TaSe$_2$ where the image is
dominated by the CDW superlattice  but somewhat similar to the situation
in 2H-NbSe$_2$ \cite{Giambattista}.

%\section{Conclusion}

The very good agreement between the real space CDW features
% namely the
% wave vectors of the various CDW modulations 
and the results from the
 neutron scattering experiments  shows that cleaved surfaces are highly ordered
and retain the electronic properties of the bulk material. A similar conclusion was
reached in ARPES experiments performed on cleaved (001)
surfaces of
\tq \cite{Zwick,Claessen}. The salient result of this work is given in fig.3
which makes it 
clear  that warming through the transverse lock-in transition the modulation evolves from
an amplitude modulation along \textbf{a} (double-$q$ superlattice) in the commensurate
phase to a phase modulation in the incommensurate wave vector regime with only a
single-$q$ vector activated over the investigated sample area. 
%%%%figure4%%%%%

\begin{figure}
\includegraphics[width=7.8cm]{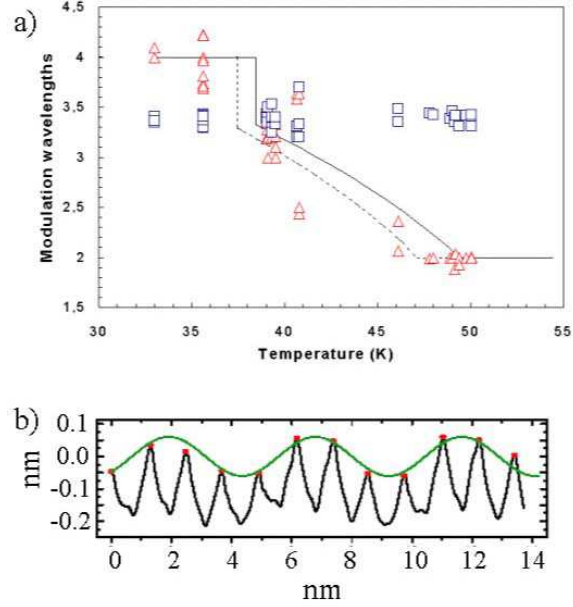}
\caption{a) Temperature dependence of the CDW wavelengths along \textbf{a}
(triangles) and \textbf{b} (open squares) directions in unit  cell dimensions.
The large scattering of the data at T=40.6K were taken from small images of 5
nm$\times$5 nm while other temperatures were taken from images larger than 10
nm$\times$10 nm. The solid (dotted) lines  are obtained from X-rays diffraction
measurements in warming (cooling) respectively. b) Cosine fit of the CDW
profile at 36.5K along the \textbf{a} direction indicated by black arrows in
fig. 3c revealing the CDW phasing.}
\label{fig.4}
\end{figure} Thus, we have shown that
\tq adopts a domain structure in the temperature regime where the transverse ordering of
the CDW's is incommensurate. This is probably the clue to understand the
hysteresis displayed by
$q_a(T)$ between 49K and 38K
\cite{Ellenson,Ellenson77} as suggested by
\cite{Pouget89,Barisic85}.

The  fact that the CDW is observable  by a STM probe shows that it 
is static in spite of its incommensurate nature
(along the \textbf{b} direction), and is therefore 
pinned by impurities or defects .

 The low temperature  CDW in TTF-TCNQ is thus an ideal candidate to
study the local phase shift for the following reasons: the unit cell in the \textbf{a-b} plane
has a quadratic symmetry, the CDW phase is commensurate in the \textbf{a} direction but
incommensurate in the \textbf{b} direction, the CDW modulation is double-$q$ modulated below
38K so the phase shifts along \textbf{a} and \textbf{b} can be studied separately and
in addition a modulation of the amplitude along
\textbf{a} is expected. 

 Furthermore, we notice on figs.4a,b that the phase of the CDW is such as to present an
alternation of the amplitude  on the \q stacks like $++--++$, etc... along the \textbf{a}
direction.

The phasing of the CDW with respect to the underlying lattice below 38 K agrees
with the  diffraction experiments data
\cite{Pouget}.  The results of our
work show that STM techniques are very well adapted to the local study of CDW's in
\tq and resolve the question of phase against amplitude modulation. 
In addition, this work opens new ways towards  a local investigation of the
pinning of the CDW's around impurities to derive information about the nature of the
pinning mechanism (strong or weak).

%\begin{acknowledgments}
We thank J.P.Pouget, K.Maki and E.Canadell for very fruitful
discussions.Z.Z.Wang acknowledges the financial support of the SESAME contract 1377.

%\end{acknowledgments}

%\bibliographystyle{unsrt}
%\bibliography{CDWa}

\end{document}